\documentclass{article}

\usepackage[english]{babel}

\usepackage[letterpaper,top=2cm,bottom=2cm,left=3cm,right=3cm,marginparwidth=1.75cm]{geometry}

\usepackage{amsmath}
\usepackage{graphicx}
\usepackage[normalem]{ulem}
\usepackage[colorlinks=true, allcolors=blue]{hyperref}
\usepackage{amsmath,amssymb,epsfig,amsfonts,mathrsfs}
\DeclareSymbolFontAlphabet{\mathrsfs}{rsfs}

\newcommand{\be}{\begin{equation}}
\newcommand{\ee}{\end{equation}}
\newcommand{\ba}{\begin{aligned}}
\newcommand{\ea}{\end{aligned}}

\title{Snowmass Whitepaper: Physical Mathematics 2021
}
\author{
Ibrahima Bah$^1$,
Daniel Freed $^2$,
Gregory W. Moore$^3$, \\
 Nikita Nekrasov$^4$,
Shlomo S. Razamat$^5$,
Sakura Sch\"afer-Nameki$^6$}

\begin{document}
\maketitle

\begin{center}
{\it $^1$  Department of Physics and Astronomy, Johns Hopkins University, Baltimore, MD 21218, USA }\\
{\it $^2$ Department of Mathematics, University of Texas at Austin }\\
{\it $^3$ NHETC and Department of Physics and Astronomy, Rutgers University }\\
{\it $^4$ Simons Center for Geometry and Physics, \\
Stony Brook University, Stony Brook, NY 11794-3636, USA}\\
{\it $^5$ Department of Physics, Technion, Haifa, 32000, Israel}\\
{\it $^6$ Mathematical Institute,University of Oxford, Oxford OX2 6GG, United Kingdom }
\end{center}
\begin{abstract}
 \noindent
This is a Snowmass whitepaper on physical mathematics. It briefly
summarizes and highlights some of the key questions drawn from
a much more extensive essay, by the same authors,
 entitled, ``A Panorama Of Physical Mathematics 2021.''  Version:    \today .
\end{abstract}


\section{General Remarks}

The history of the interactions between physics and mathematics is old
and venerable. Physics cannot flourish without mathematics and much of
mathematics takes its inspiration from physics.
The inward bound trajectory of twentieth century physics towards the
discovery of the most fundamental laws of physics resulted in the
creation of quantum field theory and string theory, hereafter abbreviated
as QFT/ST. But, while QFT/ST is a revelation
of twentieth century scientists, well into the twenty-first
century it is widely recognized as being far from  fully understood.
The implications and ramifications of QFT/ST  leave much
room for contributions by twenty-first century scientists.
The investigations into QFT/ST have
made use of ever more sophisticated mathematics, including cutting
edge mathematics at the focus of present day research. Conversely,
many developments in   QFT/ST have also
led to profound new insights, constructions, and even entire subfields of
mathematics (such as vertex operator algebra theory, or
homological mirror symmetry -- to choose but two examples, out of very many).
A community of scientists, involving both mathematicians
and physicists, is vigorously engaged in the pursuit of QFT/ST and its
relation to mathematics. In this community there has been an
important shift in viewpoint and emphasis from more traditional
deparmentalizations. There is a dual and
equal emphasis on both the discovery of the fundamental laws of nature as well
as on mathematical discovery. This field of intellectual
endeavor has, on occasion,  been called \emph{physical mathematics} and
we will adopt that term here for lack of a better designation. ``Physical mathematics'' is a
subfield of the much broader field of mathematical physics.

The essay ``A Panorama Of Physical Mathematics 2021,'' --
hereafter referred to simply as ``the Essay'' --  by
same authors as those of this whitepaper  describes
a partial snapshot, or aerial view, or panorama of the
subject as it stands in 2021. The Essay is forward-looking:
After recalling the status of the subject, some open problems
that have the potential to lead to future progress have
been identified. In what follows we briefly summarize some
of the main points from the Essay. The interested reader is urged to
consult the more extended text, where a small sampling of relevant
references may be found. Here we highlight just some of the main areas of progress,
and some promising future directions. A more comprehensive -- but still quite incomplete --
discussion can be found in the Essay.

\section{Foundational Aspects of QFT}

\subsection{Overview}

Quantum Field Theory (QFT) is the most successful tool in
existence for describing the fundamental laws of nature.
The success - in both high energy physics and in condensed matter
theory - has been nothing short of spectacular, leading
to some of the most precise, experimentally verified,
predictions in all of science. It would appear that not only is
the Book of Nature written in the language of mathematics, but that
the dialect is that of Quantum Field Theory. Nevertheless,
the vast array of phenomena QFT describes is nowhere
close to being fully fathomed. Moreover, from a mathematical
viewpoint, even giving a fully satisfying definition of
what a QFT is remains open.

Some classes of quantum field theories can be investigated
with full rigor. These include  topological quantum field theories (TQFTs),
for which there is now a well-developed mathematical theory.
Nevertheless, even some important and standard quantum field theories that are
generally regarded as TQFTs, namely, those  of ``cohomological type,''
often do not fit the rigid mathematical framework proposed by
mathematicians.
\footnote{A TQFT is said to be of \emph{cohomological type} if it is
a subsector of a metric-dependent QFT defined by taking invariants with
respect to an odd nilpotent symmetry. These are to be distinguished from
theories such as Chern-Simons theories, where the classical action is metric-independent.}
Yet the cohomological type TQFT's often lead to
some of the most interesting mathematical applications including
the discovery of new
invariants of 3- and 4-dimensional manifolds and new viewpoints
and methods in the Geometric Langlands Program.

Another important set of examples of quantum field theory are the two-dimensional
conformal field theories. At least in the case of two-dimensional rational conformal
field theories, these are mathematically completely rigorous,
thanks to the well-developed mathematics of vertex operator algebras
and the representation theory of loop groups.

Looking beyond the above examples, there are relatively new mathematical axiom systems for Wick-rotated quantum field theories on Riemannian manifolds, roughly corresponding to the Heisenberg and Schr\"odinger pictures.  Much remains to be done to develop these ideas and connect to physical theories of interest. One crucial aspect of QFT,
that mathematicians have yet to come to grips with in a fully satisfactory way, is that many QFTs have infinitely many degrees of freedom. Related to this, as has been clear from the very beginnings of the subject in the 1930's, is the need to deal with apparent infinities that arise in computations.

Other important QFTs, often those connected to self-dual theories,
are not satisfactorily defined even by the standards of textbook QFT.
Again, these theories are often bound up with some of the most interesting
new developments. These developments include
algebraic structures generalizing the theory of
vertex operator algebras, new categorical structures, connections to the
geometry of moduli spaces of general (anti-)self-dual gauge/Higgs fields,   Hitchin systems, hyperk{\"a}hler and quaternionic k{\"a}hler geometry, remarkable aspects of integrable systems, deeper insights into -- and generalizations of -- exact WKB theory, and more.
Many of these developments are connected with the formulation and discovery
of exact results that hold even
in the context of difficult, strongly coupled, interacting theories.
Often supersymmetry
is involved, and often the results involve quantities such as BPS states, that preserve some of the supersymmetry.

The full extent of the possibilities offered by
quantum field theories is mind-boggling.
Naturally, one wishes to impose some order on the apparent chaos, and
understand the laws that govern the possible laws of
nature. Ultimately one would like to understand something like the
``space of quantum field theories.'' We are a long way from doing that,
but some interesting progress on this difficult topic is being made.
Some of the issues regarding classification of the
kinds of theories that have been at the heart of physical mathematics
are addressed in the Essay. Given the depth of the mathematics involved in
TQFT and two-dimensional rational conformal field theory we have every
reason to expect that a full understanding of QFT will lead to novel
and profound mathematical structures.

\subsection{Recent Progress And Future Directions}

\paragraph{What Is A QFT?}
As we have noted, one of the most pressing challenges is to understand what,
precisely, is the definition of a QFT. No universally applicable
systematic or axiomatic definition exists.
Then, one would like to understand what properties
QFTs can have.
For example, what is a suitably general notion of a symmetry of a QFT?
Furthermore, we would like to be able to do explicit computations of the spectrum of
physically relevant operators and of correlation functions and partition functions, including in the presence of boundaries and defects.
Moreover, the connection with string theory has shown that traditional textbook views of what
constitutes a QFT vastly underestimated the collection of  QFTs which are expected to exist. The geometric approach, discussed later on, will provide some hints about how one might  develop a classification program in tandem with string theory.

\paragraph{Topological Quantum Field Theory.}

As noted above, the most mathematically developed QFT's are the
\emph{topological} quantum field theories (TQFTs).
In recent years there has been a vigorous development of this topic, and
mathematicians have extended these ideas in very sophisticated ways. There are deep applications to disparate problems in topology, symplectic geometry, higher algebra, geometric representation theory, number theory, and beyond.
Although the subject has matured,  there remain many important open questions.
For example, the study of defects and boundary conditions has much future potential, as does the development of partially defined TQFT's. (An $n$-dimensional
TQFT is said to be \emph{fully extended} or \emph{fully local} if $(n-k)$ categories can
be associated to all compact $k$-manifolds with corners. In the partially defined case
this is only possible for a certain range of $k$. Some of the most important TQFT's are in
fact only partially defined.)

TQFT's have many applications in QFT, string theory, and condensed matter theory. Motivated (in part) by these applications there has been recent progress in the theory of anomalies and the related classification of invertible field theories (including nontopological theories).  Novel anomalies constrain the dynamics of QFT's, and the topological methods arising from the classification of invertible phases are being used in general QFT as well.  Much remains to be done in these directions, and there are also
many important classification questions which persist.

\paragraph{Algebraic Structures.}

The study of the operator product expansion in QFT has
led to the discovery of many important algebraic structures.
One of the most famous is vertex operator algebra theory,
which can be traced to the operator product in two-dimensional
conformal field theory. Applying techniques of topological,
and holomorphic-topological twists to higher dimensional
theories has led to many remarkable generalizations of
vertex operator algebras. Another formalization of current
interest is the entire subject of factorization algebras.

\paragraph{Generalizations Of Symmetries.}
Another important development of the past few years is an increasingly improved understanding of
various generalizations of the notion of symmetry. One physical application of these generalizations has been a set of new insights into the   phase structure of certain strongly coupled QFTs, together with new constraints on the  IR dynamics of these theories. These applications make use of anomalies for these generalized symmetries. One generalization of global symmetries which is being intensively studied are the  higher-form symmetries, where charged objects can be associated to positive dimension submanifolds of spacetime.
The higher-form symmetries might not form a traditional group structure but rather might involve
generalizations of groups known as ``higher groups,''  or -- even more generally -- ``categorical symmetries.''

This multitude of generalized notions of symmetries are the subject of intense current scrutiny and are expected to play a fundamental role in future formulations of, and applications of, QFT.  One
notable application of higher-form symmetries has been to confinement in non-abelian Yang-Mills theory.   Another, possibly related, fruitful direction of research continues to be the study of both $\infty$-structures and BV geometry.  But the full scope of the physical implications of generalized symmetries remains to be seen, and promises to be far-reaching.

\paragraph{Integrability.}
There are several points of contact between QFTs and integrable systems.   Starting with the integrable quantum field theories in $1+1$ dimensions, a novel type of integrability emerges in the planar limit of 4d $\mathcal{N}=4$ Super-Yang Mills theory. This connection led to a vast generalization of the types of integrable systems and techniques to solve them.
Another point of contact with quantum integrable systems arises in the study of supersymmetric vacua of gauge theories with lower
supersymmetry and in various spacetime dimensions.  The search for a universal quantum algebra acting between these vacua might bring yet another generalization of the notion of symmetry in QFT.

More broadly, integrability and solvability can be used as a fundamental organizing theme
for the study of all of physical mathematics, and this viewpoint suggests an astonishing number
of new directions for research. Many of these are discussed in some detail in section 10 of
the Essay.

\section{Interactions Between Condensed Matter Physics, QFT, and Higher Mathematics}

\subsection{Overview}

An amazing aspect of physical mathematics
is that, while much of the initial impetus
was tied to string theory and high energy particle physics of the highest
energies imaginable, some of the same mathematical structures turn out to
be of use in questions involving condensed matter physics at the lowest
energies imaginable. An old example of this is elementary homotopy theory.
This turned out to be of use in discussions of solitons in QFT and that application
inspired the use of homotopy theory in describing point, line, and surface
defects in real materials. Conversely, mathematics used to describe crystals and
crystallographic groups, and even quasicrystals,
found useful applications in perturbative string compactifications
such as toroidal compactifications and orbifolds thereof.
In the past several decades there has been an explosion of
interest in the condensed matter community  in applications of topology to
condensed matter physics. For example, the (fractional) quantum Hall effect
has been an enduring source of amazing physical phenomena related
to deep mathematical constructions involving Chern classes,
Chern-Simons theory, lattice theory,
and noncommutative geometry.
Applications of modular tensor categories to the study of
two dimensional  conformal field theory
led to applications to anyon physics, quantum computing,
and quantum information theory.

\subsection{Recent Progress And Future Directions}

A more recent example of applications of
topological ideas in condensed matter physics is the study of topological insulators,
topological superconductors, and Weyl semimetals. For example the classification of
crystalline band structure insulators involves twisted equivariant
K-theory, and nontrivial K-theoretic torsion invariants associated to topological insulators
have led to striking experimental predictions,  which have been
dramatically confirmed in the laboratory. Relations between the topological invariants of
an insulator with properties of its boundary modes, when it has a boundary,
have been an important theme of research, but more needs to be done here.
An interesting question for the future is whether
other twistings of K-theory, or the related subject of ``coarse geometry,''
will find a natural home in condensed matter theory.

The study of TQFT and its relation to bordism theory has proven
to be of great use in the study of short-range entangled phases of matter and quantum information theory. Much remains to be done in this program of classifying phases of matter.  Another fertile area for investigation is the detailed relation between topological phases associated to discrete models (such as lattice models) and continuum field theories (which are possibly topological).
Recently fracton models have been under intense study since they pose
some interesting challenges in this area. These fracton models raise interesting questions
about relations of foliation theory to QFT.

While these topics constitute applications to condensed matter physics they should also be considered applications to mathematics, because the applications make use of very sophisticated mathematics and the questions inspired by the applications are in turn driving further developments in
the relevant mathematics.

\section{Low Dimensional Topology And Manifold Invariants}

\subsection{Overview}

Some of the most dramatic developments in physical mathematics
have been connected
to both differential geometry and low dimensional topology.
Here we highlight some of the most important developments.

One of the most surprising and notable developments is the entire subject of (homological) mirror symmetry in the context of enumerative algebraic geometry and symplectic topology.  This important area of mathematics has its roots in the computation of string worldsheet instanton effects in Calabi-Yau compactification of string theory. Another dramatic and important development is the relation between intersection theory on the moduli space of curves and matrix models. This development too has roots in physics. It came as a synthesis of ideas of topological field theory with the matrix model formulations of two-dimensional quantum gravity. More recently, there have been numerous physically-inspired generalizations
of  (homological) invariants of knots and links in three-manifolds, along with  new invariants of three-manifolds. Finally, in one of the paradigmatic tales of the development of physical mathematics the physical insights of instantons in gauge theory together with topological field theory have led to dramatic developments in the understanding of the differential topology of four-manifolds.

Remarkable progress has been made in Thurston's geometrization conjecture (now a theorem) for  three-manifolds. This progress has points of contact with string theory,
the renormalization group flow of sigma models, and the use of
functionals with monotonic behavior under the renormalization group.

Finally, the constructions of manifolds of special (including exceptional) holonomy, and the exploration of their moduli spaces of such structures, has seen dramatic progress, e.g. with concrete constructions of families of compact exceptional holonomy spaces.

\subsection{Recent Progress And Future Directions}

All of the above subjects continue to be a source of new ideas and progress:

The relation of matrix models to intersection theory on moduli
spaces has interesting relations to ``geometric recursion'' and ``topological
recursion,'' as well as to models of two-dimensional gravity. One would expect
that there are analogs of these relations for super-Riemann surfaces. These
are all topics of current research.

(Homological) mirror symmetry has been given rigorous proofs in special
examples but remains conjectural in many cases. The full scope of the conjecture
continues to be a source of research. The case of three-dimensional ``mirror symmetry''
is a very vigorous area of mathematical research. Via the connection with the
Geometric Langlands Program there are many important connections to geometric representation theory.

Homological invariants of knots and links have remarkable connections to
conformal field theories, Landau-Ginzburg theories, monopole moduli spaces,
integrable models, noncompact Chern-Simons theory, and more. The intricate
connections between all these topics continue to be  actively pursued.

Although the progress in Thurston's geometrization conjecture makes use
of many ideas from physics a truly satisfying synthesis of the subject with
physical models is still a topic of current research. There is still no really
workable classification of three-manifolds. It remains to be seen if physical
ideas can change the situation.

The differential topology of four-manifolds remains one of the most challenging
subjects in low dimensional topology. While supersymmetric gauge theory has led to
astonishing progress in this area important  problems, such as the 11/8 conjecture
and the smooth Poincar\'e conjecture remain open. It is natural to wonder if the physics of
supersymmetric QFT/ST will lead to yet more invariants of the diffeomorphism type of
four manifolds. It is also natural to wonder if QFT/ST can inform us of the structure
of ${\rm BDiff}(X)$ for a four-manifold $X$. Finally, there is much work in the mathematics
community on various flavors of Floer theory, but this work has drifted from the main attention of physicists, and it would be desirable to see more interaction between the physics and math communities in this subject.

The progress related to  manifolds of special holonomy and their moduli spaces in particular in the exceptional holonomy case remains in general a challenge.
These questions are closely connected to the geometrization of QFTs and is covered in the next section below.

\section{Geometrization Of QFTs}

\subsection{Overview: String Theory And M-Theory}

Like QFT, string theory and M-theory are subjects about which
enough is known to be recognized as mathematical entities which we would like to
define precisely. But a suitable definition would seem to be
a long way off. Nevertheless,
we know enough now to engage in remarkable dialogues with pure mathematics.

Even the traditional perturbative formulation of string theories involves
extremely subtle aspects of the algebraic geometry of supermanifolds, and open problems in this field remain the focus of modern research. In a different direction, some
reasonable-looking formulations of QFT and string theory turn out
to be mathematically inconsistent due to a precisely defined notion of an
anomaly. This insight of the 1970's and 1980's has blossomed
into a vigorous mathematical study of the geometrical
formulation of anomalies and its connection to invertible field theories.
 The subject is being developed for its own
sake. It also has important applications to the question of the consistency of string theory compactifications. Moreover, as noted above, there are applications to the dynamics of QFT and the classification of phases of matter.

Closely related to this, some modern ideas regarding quantum gravity
have led to new mathematical conjectures about moduli spaces of e.g. Calabi-Yau manifolds. They have also  raised some interesting
questions related to summing over topologies in quantum gravity, yielding new insights into the relation of matrix models to moduli spaces of Riemann surfaces. Some of these developments have  raised fundamental issues about what the proper meaning of holography should be. For example, should string theory be dual to a single QFT or an ensemble average of QFTs ?

\subsection{Recent Progress And Future Directions}

In addition to giving proper definitions of QFTs one would like to compute. Therefore, one would like to  develop mathematical tools to study strong-coupling regimes quantitatively. Recent years have seen dramatic progress in numerous approaches to tackling this question, many of which have an underlying connection with geometry and the subject of geometric engineering. Broadly speaking,  geometric engineering is the
realization of QFTs in the context of string compactifications on non-compact spaces with special holonomy, (after taking suitable limits).
The mathematical fruits of geometric engineering include
new enumerative invariants, as well as new differential geometric results following from holography and string compactification.

\paragraph{Classification Program Of SCFTs.}
String theory predicts the existence of superconformal field theories in 6d and 5d, which are intrinsically strongly-coupled UV fixed points -- and thus not accessible using standard perturbative QFT. In some sense string theory is an amazing laboratory in which we can construct QFTs and CFTs, whose existence would be missed in the textbook view of  QFT. This ``experimental'' construction  complements the  axiomatic approaches to defining and understanding  QFTs and CFTs.
The full classification of 6d, 5d, and 4d SCFTs is probably out of reach in the near future, but the more narrowly defined class of those which can be geometrically engineered is probably susceptible to a rigorous classification program. Even this more narrowly defined classification remains a great challenge, but one which is reasonable to pursue, and which will likely lead to rapid progress.
The realization of 6d and 5d CFTs in F-theory and  M-theory, respectively,  gives a direct connection between QFTs and so-called canonical singularities in algebraic geometry. In 6d this program has been completed, but a mathematically precise and detailed analysis of the Mori
 minimal model program in algebraic geometry, applied to canonical three-fold singularities, has yet to be carried out to achieve a classification of 5d SCFTs. Much of this interconnects with the structure of generalized symmetries and enumerative invariants, which play an essential role in characterizing the physical properties of these strongly-coupled QFTs. Another direction that has seen substantial progress in the past years is the charaterization of the (quantum) moduli spaces of theories with 8 supercharges, and their relation to so-called magnetic quivers and hyperk\"ahler singularities.

A huge remaining challenge is the classification of 4d SCFTs, in particular those with $\mathcal{N}=1$ supersymmetry. Geometrically their realization is either in terms of so-called $G_2$ manifolds in M-theory or elliptic Calabi-Yau four-folds in F-theory. In contrast to the singular Calabi-Yau three-fold geometries underlying 5d and 6d SCFTs, these geometries provide a far larger challenge in mathematics. This applies both to the Mori minimal model program, but also the challenges that exceptional holonomy spaces pose in differential geometry. Interconnected with that are the enumerative invariants, which determine the low energy effective theory of the string theory compactified on such spaces. This, in turn, opens up   challenging  questions regarding enumerative invariants related to -- for example --
the counting of   associative three-cycles in manifolds of $G_2$ holonomy.

Geometric constructions of QFTs in string theory, relate naturally with other (inter)faces of physical mathematics, e.g., as discussed below,  enumerative invariants from counting BPS states are related to analytic number theory. But they are also related to algebraic geometry and the theory of generalized symmetries. In this way physical insights can lead to  surprising interconnections within pure mathematics itself.

\paragraph{Compactification From 6d.}
Starting from 6d, compactification results in new, 
often strongly coupled, theories in lower dimensions. The Class ${\cal S}$ (and its less supersymmetric counsins) construction is the main paradigm, where the geometry of a Riemann surface determines the properties of the 4d supersymmetric QFT (SQFT). Similar connections between three-manifolds and four-manifolds and their invariants have been developed in the context of reductions to 3d and 2d. For example, the 3d-3d correspondence has led to a wealth of insights into new constructions of 3d supersymmetric field theories and their relations to 3-manifold topology.
Using the geometric constructions one can systematically understand and classify complicated strong coupling phenomena in lower dimensions such as dualities of different kinds and the emergence of symmetry. A goal of this program can be stated as an attempt to find the most general lower dimensional QFT. An example of a question one can try to address is whether any (S)CFT in dimensions lower than five can be constructed by deforming a Gaussian fixed point in the UV and whether all conceivable (S)CFTs can be found in one of these geometric setups.

\paragraph{Holography And Geometry.}  Holography is yet another central framework for exploring and engineering QFTs beyound the Lagrangian paradigm.  In holography, the classification question can be made precise and mathematically rigorous by reformulating it as the space of AdS solutions, of various dimensions, in string theory.  Such spaces can be defined as the set of solutions of some PDEs such as Monge-Amp\`{e}re equations obtained from the Einstein equations of M-theory.  An ambitious goal is to study the different classes of PDEs from Einstein equations in supergravity that can define the space of QFTs, and understand the various mathematical tools that either exist or must be developed to characterize their solution spaces.  Often results in holography provides complementary points of view to the geometric engineering perspectives above.  It is also important to characterize how various systems of PDEs encode the data of QFTs whether it is in their Lagrangian formulations, in their geometric engineering in ST or from compactifications from 6d.

\section{Relations To Number Theory}

\subsection{Overview}

Naively, one might imagine that number theory has little
to do with QFT/ST. Nevertheless, there turn out to be many
aspects of number theory that appear to be closely related
to questions in QFT/ST. One of the most common ways in which
number theoretic questions arise comes about when one studies
partition functions, or enumerates protected operators, or
BPS states. Then aspects of analytic number theory, and
in particular the theory of automorphic forms, become
quite relevant. Two examples are, first,
expressions of geometric symmetry, or duality, of (supersymmetric)
theories through the automorphic properties of their partition functions, and, second,
connections of BPS state counting functions to Poincar\'e series, Rademacher summability, and Eichler cohomology. Some of these
connections to analytic number theory have led to generalized notions
of automorphy such as mock modularity and quantum modularity, as discussed below.

There are other ways in which ideas of number theory appear to
be relevant. One example is the relation between the
attractor mechanism for the construction of supersymmetric
black holes (or flux vacua) and Calabi-Yau manifolds with  special Hodge structures. Related to this are
curious relations of the attractor mechanism, as well as generalizations
of string theory over finite fields, to some  arithmetic
aspects of Calabi-Yau manifolds. Perhaps the most compelling of the
bridges to number theory has been
the relation of supersymmetric quantum field theory with
the Geometric Langlands Program. In the past few years other
intriguing ideas for connections between number theory and QFT
have been proposed including
``arithmetic QFT'' and ``motivic QFT.''

\subsection{Recent Progress and Future Directions}

\paragraph{Enumerative Invariants.}
Counting curves, or more generally calibrated cycles, in string compactifications sometimes results in partition functions that have intriguing number theoretic properties. A good example is given by K3 surfaces. A great deal of work is being done on enumerative algebraic geometry related to physical models. For example the study of Donaldson-Thomas invariants, which count certain wrapped brane states in string theory, is a very active field in mathematics that has seen a lot of recent progress. Recent progress includes generalizations to higher-dimensional and/or non-compact Calabi-Yau spaces. Sometimes the counting functions exhibit interesting automorphic properties. In the future we can expect progress on the enumeration of other calibrated cycles in special holonomy manifolds, where the mathematical technology is still being developed. Insights from    string theory might well be crucial for future progress. It remains to be seen if these new counting functions have interesting arithmetic properties.

\paragraph{Mock Modularity And Quantum Modularity.} Important generalizations of modularity turn out to be closely connected to physical phenomena. For example mock modularity has made an appearance in the context of partition functions of topologically twisted Yang-Mills theory, related to four-manifolds, and three-manifolds.  But it has also shown up in other contexts such as the
AdS/CFT correspondence, BPS state counting, the elliptic genus of
noncompact sigma models, indices of supersymmetric gauge theories, and umbral Moonshine.
Some understanding of the origin of mock modularity (and its relation in general to noncompactness of field space) has been achieved, but a deeper understanding, that ties together all the above
instances, would be more desirable. An example of an important concrete question in this subject is the formulation of a suitable representation-theoretic interpretation of mock modularity.

Some of the new three-manifold invariants, and some BPS state counting functions
have intriguing relations to ``quantum modular forms,'' a
mathematical concept that is still under development. It is
possible that physical ideas could help guide the theory of quantum modular forms.

\paragraph{Geometric Langlands Program.} As mentioned above,
one the deepest and most intensively studied connections
to number theory is the large body of work related to the
geometric Langlands program and its connection to boundaries,
defects, and duality symmetries of supersymmetric Yang-Mills.
While much studied, many important open questions remain
and are being actively pursued. The Essay describes several of
these questions and potentially fruitful directions these
questions have inspired.

\section{Many Omissions}

The authors of the Essay have made some effort to cast a wide net,
but many of the big ones got away. Even the broadest panorama will
miss shining wonders that lie beyond the observer's horizon.
Three examples are, fortunately, covered by other Snowmass documents.
The first example is the splendid topic of Moonshine.  The second is
the collection of remarkable mathematical developments
in the theory of perturbative QFT/ST amplitudes. The third is the
important and vast topic of mathematical aspects of general relativity.
The list of topics which
\underline{should} have been
covered in the Essay -- but are not --  remains long, and surely contains some jewels
that will be of great importance to future generations.

\subsection*{Acknowledgements}


We would like to thank  David Ben-Zvi, Lakshya Bhardwaj, Daniel Brennan, Nikolay Bobev, Miranda Cheng, Mykola Dedushenko,
Emanuel Diaconescu, Tudor Dimofte, Lance Dixon, Yakov Eliashberg, Michael Freedman, Arthur Jaffe, Jeff Harvey, Mike Hopkins,
Dominic Joyce, Anton Kapustin, Ahsan Khan, Alexei Kitaev, Zohar Komargodski, Maxim Kontsevich, Craig Lawrie, Andrei Losev, Marcos Marino, Dave Morrison, Andy Neitzke, Natalie Paquette, Sara Pasquetti, Peter E. Pushkar, Pavel
Safronov, Nathan Seiberg, Shu-Heng Shao, Yuji Tachikawa, Constantin Teleman, Thomas Walpuski,
Maxim Zabzine for useful correspondence and discussions.

The  work  of  IB  is  supported  in  part  by  NSF  grant  PHY-2112699  and  in  part  by  the  “Simons Collaboration on Global Categorical Symmetry”.
DF is partially supported by the
National Science Foundation under Grant Number DMS-2005286 and partially by the ``Simons Collaboration on Global Categorical Symmetries".
The work of GM is supported by US Department of Energy under grant DE-SC0010008.
Part of the work of NN was done while he was a visiting professor
at the CAS (Skoltech), and Kharkevich IPPI in Moscow, Russia, as well as visiting the Isla San Francisco in Puccalpa, Peru. He thanks these (non-government, public-sponsored) institutions for hospitality and support. SSN is supported in part by the European Union's Horizon 2020 Framework: ERC grant 682608 and in part by the ``Simons Collaboration on Special Holonomy in Geometry, Analysis and Physics''. The research of SSR was supported in part by Israel Science Foundation under grant no. 2289/18, by a Grant No. I-1515-303./2019 from the GIF, the German-Israeli Foundation for Scientific Research and Development, by BSF grant no. 2018204.

\end{document}